\documentclass[a4paper,12pt]{article}

\usepackage[english]{babel}
\usepackage[utf8x]{inputenc}
\usepackage[T1]{fontenc}

\usepackage[a4paper,top=3cm,bottom=2cm,left=3cm,right=3cm,marginparwidth=2cm]{geometry}

\usepackage{setspace}
\usepackage{amssymb}

\usepackage{cite}

\usepackage{amsmath}
\usepackage{graphicx}
\usepackage[colorinlistoftodos]{todonotes}
\usepackage[colorlinks=true, allcolors=blue]{hyperref}

\title{Broadband switchable terahertz half-/quarter-wave plate based on a graphene-metal hybrid metasurface}

\author{Xiaoqing Luo$^{1}$, Juan Luo$^{1,2}$,  Fangrong Hu$^{1}$, and Guangyuan Li$^{1,*}$}

\date{}
\begin{document}
\maketitle

\begin{spacing}{2.0}

\noindent \large$^1$CAS Key Laboratory of Human-Machine Intelligence-Synergy Systems, Shenzhen Institute of Advanced Technology, Chinese Academy of Sciences, Shenzhen 518055, China

\noindent $^2$Guangxi Key Laboratory of Optoelectronic Information Processing, Guilin University of Electronic Technology, Guilin 541004, China


\noindent *Corresponding author: gy.li@siat.ac.cn

\end{spacing}

\newpage

\begin{abstract}
Metasurfaces incorporating graphene hold great promise for dynamic manipulation of terahertz waves. However, it remains challenging to design a broadband graphene-based terahertz metasurface with switchable functionality of half-wave plate (HWP) and quarter-wave plate (QWP). Here, we propose a graphene-metal hybrid metasurface for achieving broadband switchable HWP/QWP in the terahertz regime. Simulation results show that, by varying the Fermi energy of graphene from 0~eV to 1~eV, the function of the reflective metasurface can be switched from an HWP with polarization conversion ratio exceeding 97\% over a wide band ranging from 0.7~THz to 1.3~THz, to a QWP with ellipticity above 0.92 over 0.78--1.33~THz. The sharing bandwidth reaches up to 0.52~THz and the relative bandwidth is as high as 50\%. We expect this broadband and dynamically switchable terahertz HWP/QWP will find applications in terahertz sensing, imaging, and telecommunications.
\end{abstract}

\newpage

\section{Introduction}
Metamaterials and metasurfaces have recently emerged as promising platforms for manipulating the polarization state of electromagnetic waves because of their compactness, flexibility, and easy integration \cite{FPC2010PolMM_rev,RPP2016Metasurface_Rev,IEEEPJ2018LPC,IEEEAccess2019dual,JPD2020broadband}. Half-wave plates (HWPs) and quarter-wave plates (QWPs) are two key devices for realizing polarization conversion among two orthogonal polarization states, and the left- and right-handed circular polarization (LCP and RCP) states. In the terahertz regime, metamaterial- or metasurface-based HWPs/QWPs are of particular interest because conventional approaches based on birefringence or total internal reflection effects are usually bulky, narrowband, and sometimes lossy. 

Over the years, metamaterial-based terahertz polarization converters have evolved from narrow-band \cite{NJP2012THzMMPol,APL2013THzMMPol} to broadband \cite{Science2013HWP-THz,APL2014HWP-THZ,LPR2014QWP-THZ,OL2016Gbroadband,OE2017Gbroadband,EPL2017HWPQWPBroad}. However, the functionalities of these devices, which are made of metals or dielectrics, cannot be dynamically tuned. Recently, dynamically switchable HWPs/QWPs based on metamaterials or metasurfaces incorporating tunable materials, such as liquid crystal, vanadium dioxide (VO$_2$), or graphene, have attracted increasing attentions. Based on metamaterials incorporating liquid crystals, Vasi{\'c} \cite{Nanotech2017TunableLC} and Ji {\sl et al.} \cite{OE2018SwitchQWP} respectively proposed electrically tunable terahertz polarization converters, of which the bandwidths are only 0.056~THz and 0.35~THz, respectively. Based on VO$_2$-metal metasurfaces, Wang {\sl et al.} \cite{SCIREP2015QWPVO2,IEEEPJ2016QWPVO2} demonstrated terahertz QWPs with switchable single or multiple operation frequencies. Nakata {\sl et al.} \cite{PRA2019QWP-THz} demonstrated a switchable QWP operating at the specific frequency of 0.617~THz by designing a VO$_2$-based anisotropic checkerboard metasurface. Zhao {\sl et al.} \cite{CPL2020QWP2HWP} proposed a switchable terahertz metamaterial that can be switched between an HWP and a QWP in the spectral band of 2.09--2.27~THz, corresponding to a relatively narrow bandwidth of 0.18~THz. Quite recently, some of the authors \cite{OE2020HWP2QWP} proposed a metal-VO$_2$ metamaterial for achieving broadband switchable terahertz HWP/QWP, the bandwidth of which covers 0.66--1.40 THz, corresponding to a relative bandwidth of 71.8\%. We further proposed a novel design philosophy making use of the transition from the overdamped to the underdamped resonance, and designed a VO$_2$-metal hybrid metasurface for achieving broadband dynamically switchable HWP/QWP in the spectral band of 0.8--1.2~THz  \cite{arXiv2021HWPQWPBroad}.

Compared with liquid crystals and VO$_2$, graphene have many unique characteristics, including ultra-thin thickness (only 0.33~nm), low loss, and continuous and flexible tunability via electric biasing \cite{ACSN2012Gtune}, chemical doping \cite{JMC2011Gdoping}, or optical pumping \cite{JAP2007GOpump}. Zhang {\sl et al.} \cite{OE2015QWP2Linear-THZ} and Tavakol {\sl et al.} \cite{PTL2019TunableQWP} respectively proposed switchable QWPs based on graphene metamaterials, and showed that the polarization state of the output wave can be dynamically switched among linear, left- and right-handed polarization by changing the graphene chemical potential. Zhang {\sl et al.} \cite{OSAC2018QWP2HWP} proposed functional switch from a QWP to an HWP within 4.80--5.10~THz based on a graphene metasurface. The corresponding bandwidth is only 0.3~THz and the relative bandwidth is only 6\%. Guan {\sl et al.} \cite{OL2019HWP2QWP_GSi} propose a hybrid graphene-dielectric metasurface for realizing switchable HWP/QWP operating at the single frequency of 1~THz. Qi {\sl et al.} \cite{OE2020HWPQWPG} also proposed a graphene-based high-efficiency switchable HWP/QWP operating at the specific frequency of 15.96~THz. Quite recently, Zhang {\sl et al.} \cite{OE2020HWPQWPGbroad} proposed a hybrid graphene-metal metasurface for achieving terahertz HWP/QWP in the frequency range of 1.38--1.72~THz. However, the corresponding bandwidth is 0.34~THz and the relative bandwidth is only 22\%. Therefore, it remains challenging to design broadband switchable QWPs/HWPs based on graphene metasurfaces, greatly hindering their practical applications.

In this work, we propose a graphene-metal hybrid metasurface for achieving broadband switchable terahertz HWPs/QWPs. The metasurface unit cell is composed of a gold stripe and a graphene stripe on a thick gold film sandwiched by a dielectric spacer. We will show that the function of the proposed metasurface can be switched from a broadband HWP with polarization conversion ratio (PCR) exceeding 97\% in the band of 0.7--1.3~THz, to a broadband QWP with ellipticity over 0.92 in the range of 0.78--1.33~THz. Therefore, these two switchable functionalities share the same spectral band of 0.78--1.3~THz, corresponding to a strikingly broad bandwidth of 0.52~THz and relative bandwidth of 50\%, which is much larger than those of graphene-based metasurface HWPs/QWPs in the literature \cite{OSAC2018QWP2HWP,OL2019HWP2QWP_GSi,OE2020HWPQWPG,OE2020HWPQWPGbroad}.

\section{Simulation setup}
\begin{figure}[htp]
\centering
\includegraphics[width=\linewidth]{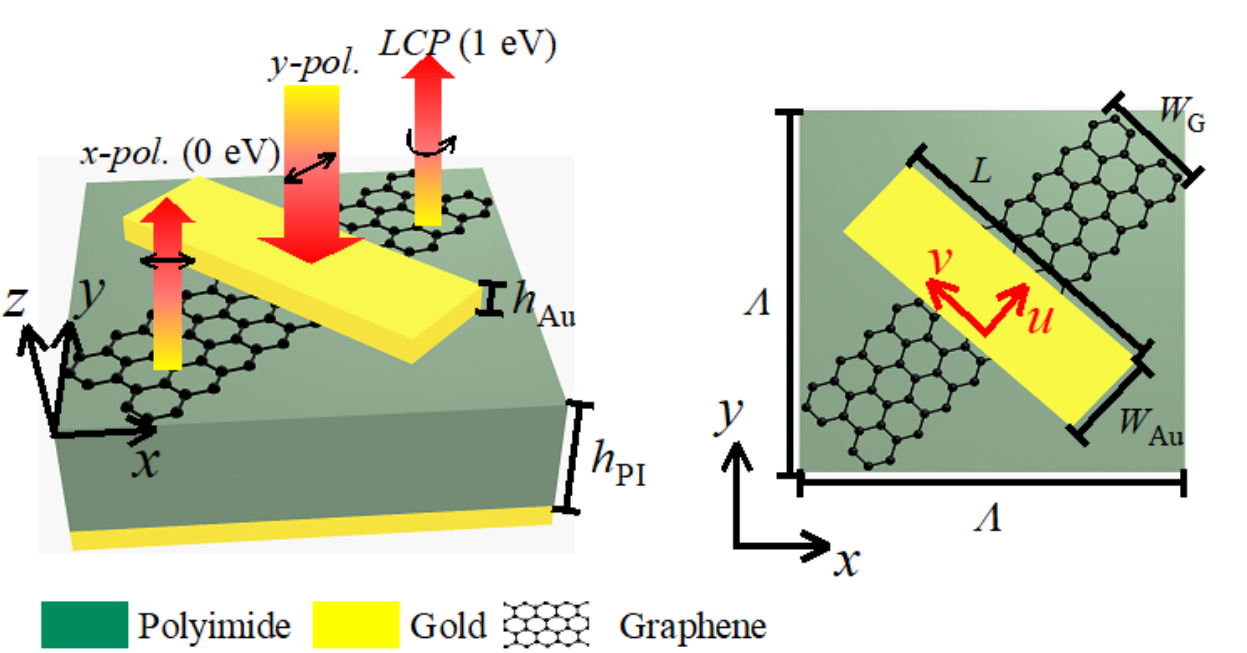}
\caption{Schematics of the proposed graphene-metal hybrid metasurface. (a) 3D view of the metasurface, which can function as an HWP converting the incident linearly $y$-polarized terahertz wave into $x$-polarized wave when the Fermi level of graphene $E_{\rm F}$ is 0~eV, or as a QWP converting the incident $y$-polarized wave into right-handed circularly polarized wave when $E_{\rm F}=1$~eV. (b) Top view of the unit cell with $\Lambda=120~\mu$m, $W_{\rm Au}=30~\mu$m, $W_{\rm G}=20~\mu$m, $L=90~\mu$m, $h_{\rm PI}=38~\mu$m, and $h_{\rm Au}=200~$nm.}
\label{fig:schem}
\end{figure}

Figure~\ref{fig:schem} illustrates the proposed metasurface acting as a broadband and dynamically switchable terahertz HWP/QWP. The metasurface unit cell consists of a gold short stripe and a graphene long stripe, which are perpendicular to each other and placed obliquely of 45$^\circ$ with respect to the $x$-axis. These graphene-gold hybrid stripes stand on a thick gold film sandwiched by a polyimide spacer, forming a metal-insulator-metal configuration. The metasurface can be fabricated using the state-of-the-art top-down microfabrication processes. A thick gold film is first deposited onto a silicon substrate, followed by spin-coating of a polyimide film of designed thickness. A monolayer graphene layer is then transferred on the top, and patterned into two-dimensional array of stripes using photolithography and reactive ion etching. Finally, the gold stripe array can be fabricated via photolithography, thermal evaporation, and liftoff processes.

The operation principle of the proposed metasurface is as follows. When the Fermi level of graphene $E_{\rm F}$ is pinned at 0~eV, the graphene conductivity is as low as the dielectric material for the terahertz wave. In this scenario, the metasurface unit cell is effectively composed of the gold stripe, and thus the metasurface works as an HWP which can convert the incident linear polarization state into the orthogonal linear polarization state. When the graphene Fermi level is tuned to $E_{\rm F}=1$~eV, which can be done via external stimulus, such as electric biasing \cite{ACSN2012Gtune} or optical pumping \cite{JAP2007GOpump}, the graphene conductivity is as high as metal-like material. In this case, the metasurface acts as a QWP that converts the linearly polarized incident wave into circularly polarized wave.

The polarization-dependent reflection amplitude and phase spectra of the proposed metasurface were numerically simulated using the frequency-domain solver in CST Microwave Studio. The unit cell boundary conditions were applied in the $x$ and $y$ directions, and the open boundary condition was used in the $z$ direction. Tetrahedral meshes with adaptive refinement process were applied. In all the simulations, gold was modelled using the lossy metal model with conductivity of $4.56 \times 10^7$~S/m, and the built-in material permittivity was adopted for polyimide. 

The surface conductivity of graphene $\sigma_{\rm G}$ can be described by the Kubo equation with intraband and interband contributions~\cite{GraphenePerformance2014}, that is
\begin{equation}
\sigma_{\rm G}=\sigma_{\rm intra}+\sigma_{\rm inter}\,.    \label{eq:sigmaG}
\end{equation}
According to Pauli exclusion principle, the interband contribution can be safely neglected compared with the intraband contribution for low terahertz frequencies at room temperature~\cite{Graphene2007}. Thus the expression of the graphene conductivity can be simplified into~\cite{GraphenePlasmonics2011}
\begin{equation}
\sigma_{\rm G} \approx \sigma_{\rm intra}=\frac{-i e^{2}k_{\rm B}T}{\pi\hbar^{2}(\omega+2i\tau)}\left[\frac{E_{\rm F}}{k_{\rm B}T}+2 \ln(e^{\frac{E_{\rm F}}{k_{\rm B}T}}+1)\right]\,.
\label{eq:sigmaG2}
\end{equation}
Here $e$ is the charge of an electron, $k_{\rm B}$ is the Boltzmann constant, $T$ is the temperature, $\omega$ is the light frequency, $\tau$ is the carrier relaxation time from the impurities in graphene, and $\hbar$ is the Planck constant.

\section{Results and discussion}
\subsection{HWP function}
\begin{figure}[htp]
\centering
\includegraphics[width=\linewidth]{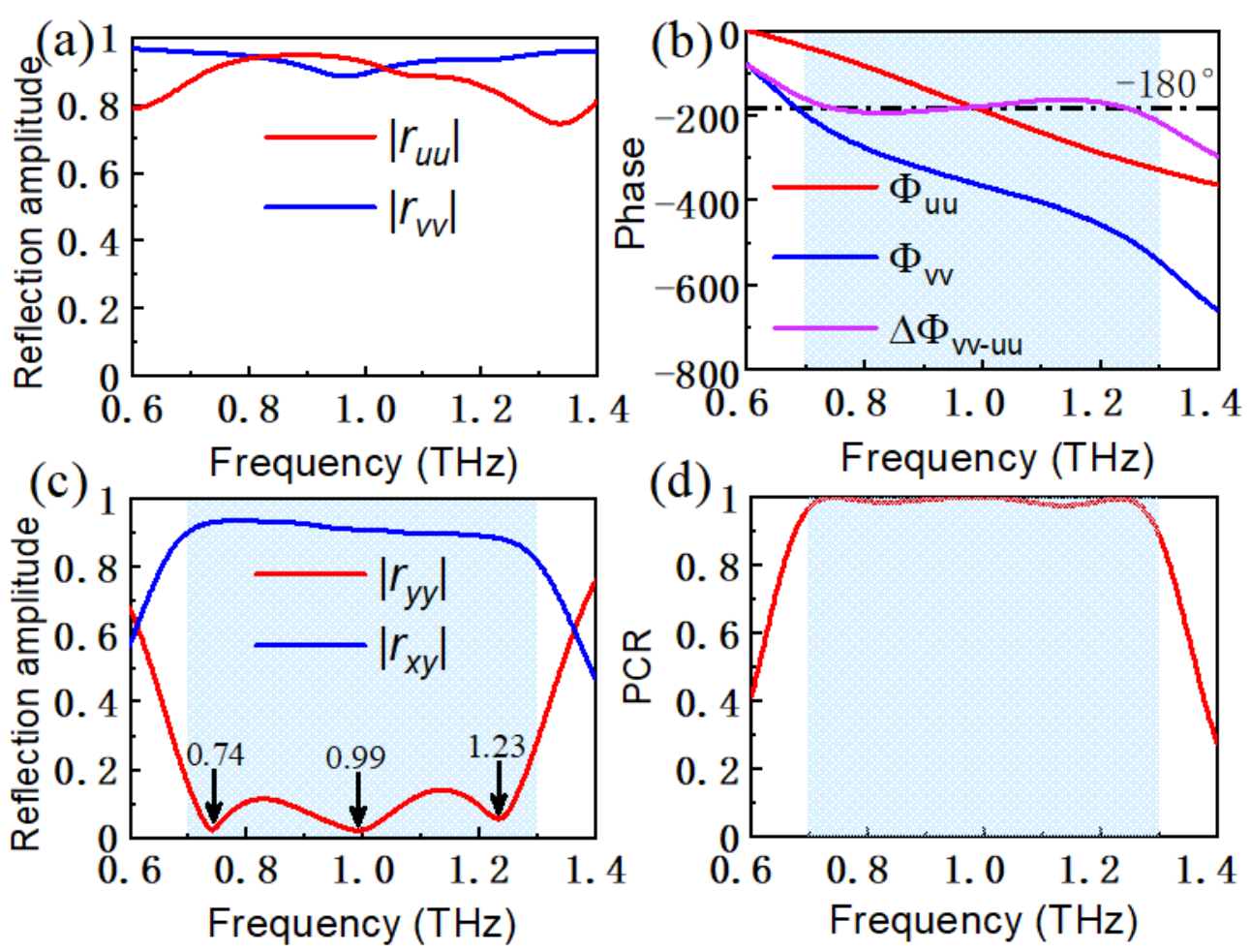}
\caption{Simulated reflection (a) amplitude $|r|$ (b) phase $\Phi$ spectra of the proposed hybrid metasurface with $E_{\rm F}=0$~eV under $u$- and $v$-polarized incidences. (c) Calculated reflection amplitude spectra for the co-polarization $|r_{yy}|$ and cross-polarization $|r_{xy}|$. (d) Calculated PCR spectra under $y$-polarized incidence. Blue regions in (b)--(d) indicate the broad operation band of the obtained HWP.}
\label{fig:SpecHWP}
\end{figure}

Figures~\ref{fig:SpecHWP}(a)(b) depict the simulated reflection amplitude and phase spectra of the proposed metasurface when the graphene Fermi level is set to be 0~eV, and under the incidence of $u$- or $v$-polarized terahertz wave. Results show that $|r_{uu}|$ and $|r_{vv}|$ are almost equal to each other, and are close to 0.9 within the frequency range of 0.7--1.3~THz. Meanwhile, their phase differences are approximately equal to $-180^{\circ}$, {\sl i.e.}, $\Delta\Phi=\Phi_{vv}-\phi_{uu}=-180^{\circ}$. Therefore, the graphene-metal hybrid metasurface with $E_{\rm F}=0$~eV can act as a broadband and efficient HWP. 

\begin{figure*}[hbtp]
\centering
\includegraphics[width=14cm]{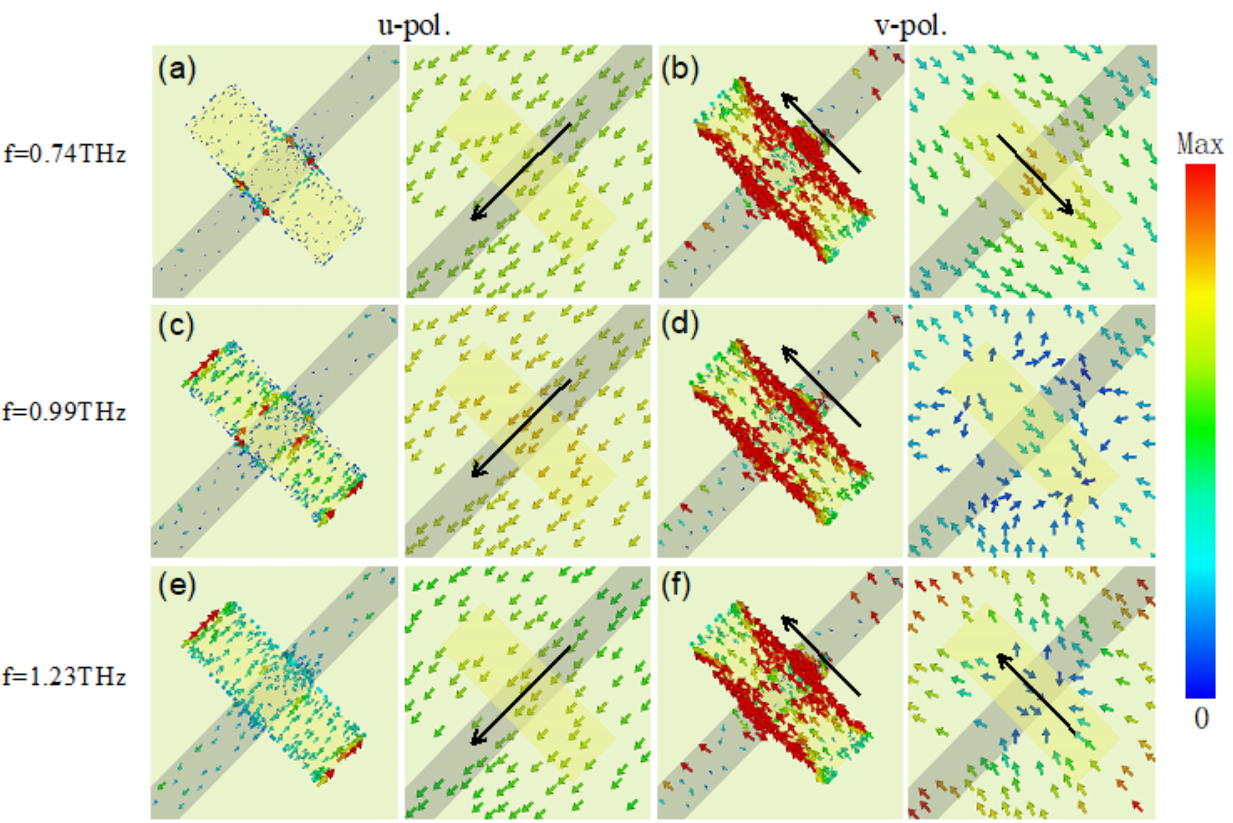}
\caption{Surface current distributions (arrows for directions and colors for strengths) on the gold and graphene stripes (1$^{\rm st}$ and 3$^{\rm rd}$ columns) and on the gold film (2$^{\rm nd}$ and 4$^{\rm th}$ columns) for three resonant frequencies of (a)(b) 0.74~THz, (c)(d) 0.99~THz, (e)(f) 1.23~THz when the graphene Fermi level is $E_{\rm F}=0$~eV. The 1$^{\rm st}$ and 2$^{\rm nd}$ columns are for the $u$-polarized incidence, and the 3$^{\rm rd}$ and 4$^{\rm th}$ columns are for the $v$-polarized incidence. The black arrows indicate the dominant current directions.}
\label{fig:FieldHWP}
\end{figure*}

In order to evaluate the performance of the obtained HWP under the $y$-polarized incidence, we calculate the co-polarized and cross-polarized reflection amplitude spectra, $|r_{yy}|$ and $|r_{xy}|$, and the spectra of the polarization conversion ratio (PCR), which can be calculated by
\begin{eqnarray}
{\rm PCR} & = & |r_{xy}|^2/(|r_{xy}|^2+|r_{yy}|^2)\,.
\label{Eq:PCR}
\end{eqnarray}
Figure~\ref{fig:SpecHWP}(c) shows that the cross-polarized reflection amplitude $|r_{xy}|$ is larger than 0.9, whereas the co-polarized reflection amplitude $|r_{yy}|$ is smaller than 0.1 in the range of 0.7--1.3~THz. For $|r_{yy}|$, there exist three dips locating at 0.74~THz, 0.99~THz and 1.23~THz. As a result, the calculated PCRs are larger than 0.97 within the broad spectral band of 0.7--1.3~THz. This corresponds to a relative bandwidth of $\Delta f/f_0 = 60\%$ with the central frequency locating at $f_0=1$~THz. These results suggest that the incident $y$-polarized terahertz wave can be efficiently converted to the $x$-polarized wave by the reflective metasurface with a broad operation bandwidth and high polarization conversion ratios. In other words, the proposed metasurface can work as a broadband and efficient HWP if the graphene Fermi level is 0~eV.

The broadband performance of the obtained HWP can be understood from the surface current distributions on the gold and graphene stripes and on the bottom gold film. Figure~\ref{fig:FieldHWP} depicts the surface current distributions under the $u$- and $v$-polarized incidences at the three resonant frequencies of $f=0.74$~THz, 0.99~THz, and 1.23~THz, for which $|r_{yy}|$ shows dips, as shown in Figure~\ref{fig:SpecHWP}(c). Figures~\ref{fig:FieldHWP}(a)(b) show that at 0.74~THz, the surface currents on the gold stripe are weak under the $u$-polarized incidence, but strong under the $v$-polarized incidence. These strong surface currents have opposite direction compared to those on the bottom gold film, producing a magnetic resonance under the $v$-polarized incidence. Similarly, at 0.99~THz and 1.23~THz, Figures~\ref{fig:FieldHWP}(c)(e) show that magnetic resonances are also generated under the $u$-polarized incidence, because the surface currents on the gold stripe have opposite direction to those on the gold film. Figure~\ref{fig:FieldHWP}(d) shows that for the $v$-polarized incidence at 0.99~THz, the surface currents on the gold stripe are strong and have clear flow direction, whereas those on the gold film have counter-propagating directions and almost cancel each other. At 1.23~THz, the surface currents on the gold stripe and on the gold film are parallel to each other, resulting in an equivalent electric resonance, as shown by Figure~\ref{fig:FieldHWP}(f). Therefore, the broadband performance of the obtained HWP should originate from the superposition of these multiple resonances.

\subsection{QWP function}
\begin{figure}[htp]
\centering
\includegraphics[width=\linewidth]{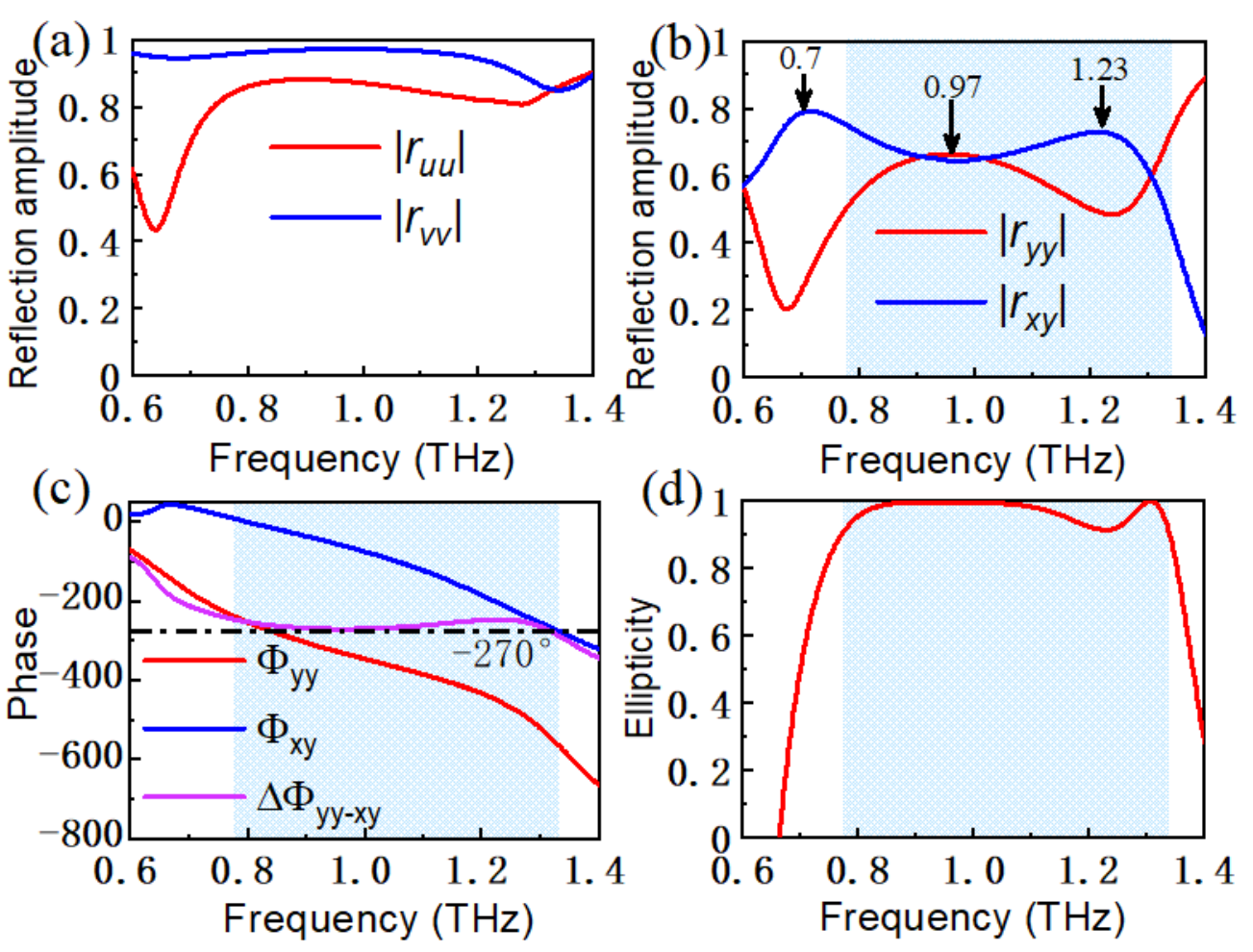}
\caption{Simulated reflection (a)(b) amplitude $|r|$ and (c) phase spectra of the proposed hybrid metasurface with $E_{\rm F}=1$~eV. (d) The calculated ellipticity under linearly $y$-polarized incidence. Blue regions in (b)--(d) indicate the broad operation band of the obtained QWP.}
\label{fig:SpecQWP}
\end{figure}

We now tune the graphene Fermi level to $E_{\rm F}=1$~eV. Figure~\ref{fig:SpecQWP}(a) shows that the reflection amplitudes $|r_{uu}|$ and $|r_{vv}|$ are close to 0.9 for frequencies above 0.80~THz. Therefore, $|r_{yy}|$ and $|r_{xy}|$ are also close to each other, and their phase differences are approximate to $-270^\circ$ within the broadband frequency range of 0.78--1.33~THz, as shown in Figures~\ref{fig:SpecQWP}(b)(c). These results suggest that the proposed metasurface with $E_{\rm F}=1$~eV now acts as a terahertz QWP.

The performance of the obtained QWP can be quantified using the ellipticity defined as $\chi \equiv S_3/S_0$, where the Stokes parameters $S_0$ and $S_3$ are expressed as \cite{LPR2014QWP-THZ},
\begin{eqnarray}\label{Eq:S}
S_0 &=& |r_{xy}|^2+|r_{yy}|^2\,,\\
S_3 &=& 2|r_{xy}||r_{yy}|\sin(\Delta\Phi)\,. 
\end{eqnarray}
Here $\Delta\Phi=\Delta\Phi_{yy}-\Delta\Phi_{xy}$ means the phase difference between co-polarization and cross-polarization reflections. When $\chi$ equals to 1 or $-1$, the polarization state of the reflective terahertz wave is LCP or RCP, respectively. Figure~\ref{fig:SpecQWP}(d) shows that $\chi>0.92$ over the broad frequency range of 0.78--1.33~THz. This result means that the graphene-metal hybrid metasurface with $E_{\rm F}=1$~eV now functions as a broadband and efficient QWP that converts the incident linear $y$ polarization into the RCP.

\begin{figure*}[htp]
\centering
\includegraphics[width=14cm]{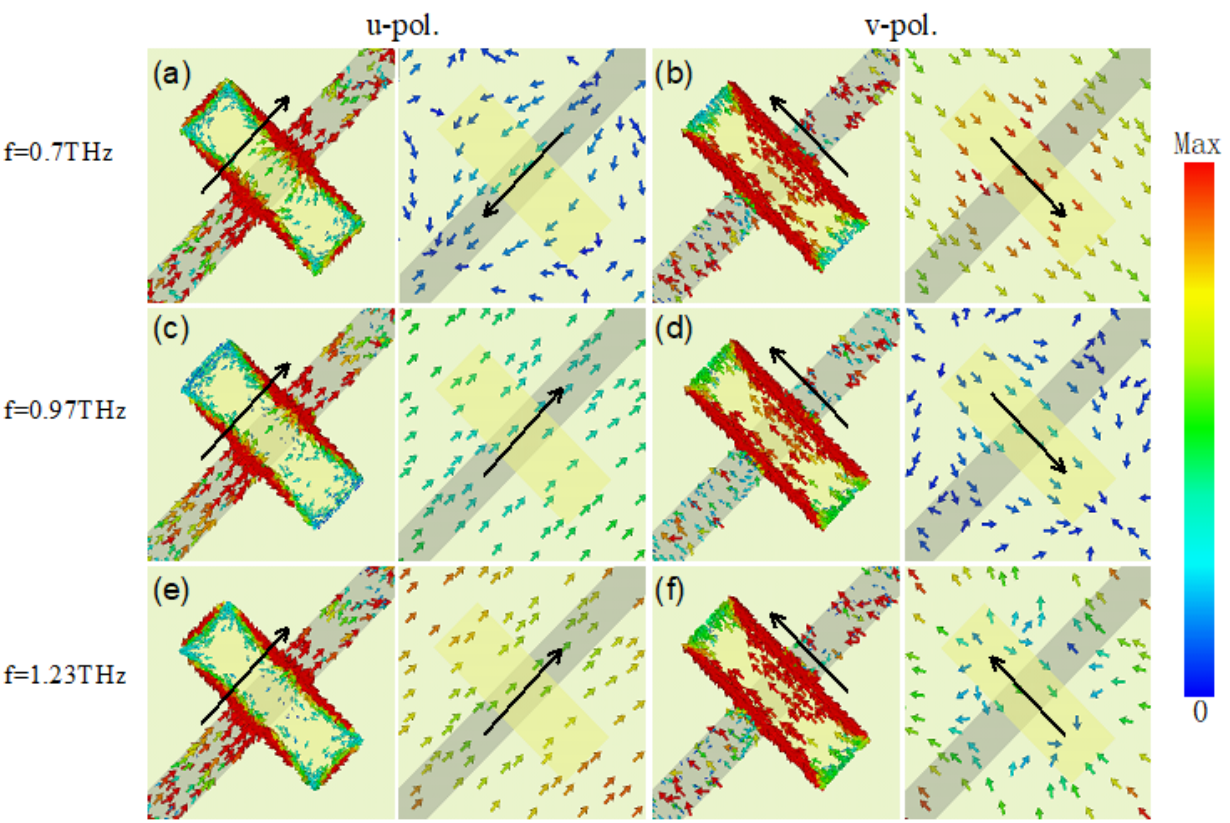}
\caption{Similar to Figure~\ref{fig:FieldHWP} but for (a)(b) 0.7~THz, (c)(d)0.97~THz, and (e)(f) 1.23~THz when $E_{\rm F}=1$~eV.}
\label{fig:CurrentQWP}
\end{figure*}

To understand the broadband performance of the obtained QWP, we also plot the surface current distributions on the gold and graphene stripes and on the bottom gold film. Here we use the three frequencies of 0.7~THz, 1.23~THz and 0.97~THz, which corresponds to two peaks and a dip in the $|r_{xy}|$ spectra, respectively, as shown by Figure~\ref{fig:SpecQWP}(b). Figures~\ref{fig:CurrentQWP}(a)(b) show that at 0.7~THz, the surface currents on the gold and graphene stripes have the opposite direction compared with those on the gold film, forming magnetic resonances for both the $u$-polarized and $v$-polarized incidences. For the $u$-polarized incidence at 0.97~THz, and for both the $u$-polarized and $v$-polarized incidences at 1.23~THz, the surface currents the gold and graphene stripes and on the gold film have the same directions, resulting equivalent electric resonances, as shown by Figures~\ref{fig:CurrentQWP}(c)(e)(f). For the $v$-polarized incidence at 0.97~THz, Figure~\ref{fig:CurrentQWP}(d) shows that the surface currents on the gold and graphene stripes are strong and have well defined direction, whereas those on the gold film are weak and have counter-propagating directions. Therefore, the broadband performance of the obtained QWP should also originate from the superposition of multiple resonances.

\section{Conclusions}
In conclusion, we have proposed a broadband switchable terahertz HWP/QWP based on a graphene-metal hybrid metasurface. Simulation results have shown that, by varying the Fermi energy of graphene from 0~eV to 1~eV, the function of the reflective metasurface can be switched from an HWP, which converts the incident $y$-polarized terahertz wave into $x$-polarized wave, to a QWP that converts the $y$-polarized wave into right-handed circularly polarized wave. The polarization conversion ratio of the obtained HWP exceeds 97\% over a wide band ranging from 0.7~THz to 1.3~THz, and the ellipticity of the obtained QWP is larger than 0.92 over 0.78--1.33~THz. Therefore, we have shown that the proposed metasurface can function either as an HWP or as a QWP in the spectral band of 0.78--1.3~THz. This corresponds to a strikingly broad bandwidth of 0.52~THz and relative bandwidth of 50\%, which is much larger than the literature on graphene-based metasurface HWPs/QWPs. The broadband properties have been explained with the superposition of multiple resonances. We expect the designed broadband and dynamically switchable terahertz HWP/QWP will find applications in terahertz polarization-dependent systems of sensing, imaging, and telecommunications.


\section*{Acknowledgments}
This work was supported by the Shenzhen Research Foundation (JCYJ20180507182444250).

\bibliographystyle{unsrt}
\bibliography{sample}

\end{document}